\documentclass[preprint,12pt]{article}
\usepackage{fullpage}

\usepackage[textsize=tiny,color=green]{todonotes}
\usepackage{url} 
\usepackage{soul}
\usepackage{color,soul}
\usepackage{hyperref}
\usepackage{paralist}
\usepackage{threeparttable}
\usepackage{comment}

\usepackage{titling}

\specialcomment{critis}{\begingroup\color{brow}}{\endgroup}
\specialcomment{asia}{\begingroup\color{red}}{\endgroup}
\specialcomment{pst}{\begingroup\color{gray}}{\endgroup}

\specialcomment{old}{\begingroup\color{blue}}{\endgroup}
\specialcomment{progress}{\begingroup\color{blue}}{\endgroup}

\excludecomment{pst}
\excludecomment{old}
\excludecomment{asia}
\excludecomment{critis}
\excludecomment{progress}

\newcommand{\xcite}[1]{}
\newcommand{\U}{USBCaptchaIn}

\begin{document}



\title{ USBCaptchaIn: Preventing \big(Un\big)Conventional Attacks from Promiscuously Used USB Devices in Industrial Control Systems. }

\author{Federico Griscioli and Maurizio Pizzonia\\
Universit\`a degli Studi Roma Tre, Engineering Dept., CS and Aut. Sec.\\Via della Vasca Navale 79, 00146 Roma, Italy}
\date{}
\newcommand{\Addresses}{{
		\footnotesize
		\textsc{Universit\`a degli Studi Roma Tre, Engineering Dept., CS and Aut. Sec.\\Via della Vasca Navale 79, 00146 Roma, Italy}
}}
\maketitle


\begin{abstract}
	
Industrial Control Systems (ICS) are sensible targets for high profile attackers
and advanced persistent threats, which are known to exploit USB thumb drives as
an effective spreading vector. 
In ICSes, thumb drives are widely used to transfer files among disconnected
systems and represent a serious security risks, since, they may be promiscuously
used in both critical and regular systems. 
The threats come both from malware hidden in files stored in the thumb drives
and from BadUSB attacks~\cite{badusblackhat2014}. 
BadUSB leverages the modification of firmware of USB devices in order to mimic
the behaviour of a keyboard and send malicious commands to the host.
%

We present a solution that allows a promiscuous use of USB thumbs drives while
protecting critical machines from malware,
that spread by regular file infection or by firmware infection.
The main component of the architecture we propose is an hardware, called \U, 
intended to be in the middle between a critical machine and all USB devices.
We do not require users to change the way they use thumb drives. To
avoid human-errors, we do not require users to take any decision. The proposed approach is
highly compatible with already deployed products of a ICS environment
and proactively blocks malware before they reach their targets. We
describe our solution, provide a thorough
analysis of the security of our approach in the ICS context, and report 
the informal feedback of some experts regarding our first prototypes.

\end{abstract}




\section{Introduction}\label{sec:intr}


Cyber-attacks to critical infrastructures constitute a serious risk for society~\cite{lewis2014critical}.
%
Specifically crafted malware can be used by attackers to alter an industrial
process or gather industrial secrets, and, in the end, gain some market or
political advantage.
%
%
Due to the inherent criticality of ICSes, best
practices~\cite{stouffer2011guide} suggest to isolate the most critical parts of
the system from other IT components, either physically or by means of firewalls.
To overcome the limitation of a poorly connected environment, file transfers are
usually performed by means of USB thumb drives and other \emph{Removable Storage
	Devices} (\emph{RSD}es). The use of RSDes turned out to be an important vector
of malware spread~\cite{rautmare2011scada} making isolation efforts to protect
ICSes from the rest of the IT systems largely ineffective.
Further, the recent class of attacks called BadUSB~\cite{badusblackhat2014} disclosed new 
threats that involve USB devices. 
These attacks are based on a modification of the device firmware, that forces
the infected device, typically a thumb drive, to behave as a different kind of
device, for example as a keyboard. In this way, a \emph{malicious firmware} can 
inject commands that end up in a malware infecting the
host. These malicious commands could download a malware from the Internet, but
can also create it on-the-fly, for example, by ``typing'' the content of a
malicious script and running it.

Regular antiviruses are largely ineffective against innovative malware, i.e.
malware that exploit zero-day vulnerabilities, and, especially against BadUSB
attacks since they exploit basic capabilities (e.g., typing operating system
commands) that the user is normally allowed to use.

In our approach,  machines are either \emph{critical} (SCADA, embedded devices, etc.)
or \emph{regular} (personal notebooks, company PC, etc.). We consider regular machines
and RSDes as possible sources and vectors of attacks against critical machines.

In this paper, we propose a solution that enables promiscuous use of
RSDes in critical infrastructures, while preventing the spread of both
conventional and firmware-based malware into the critical machines.

\begin{figure*}
	\centering
	\includegraphics[width=1\columnwidth]{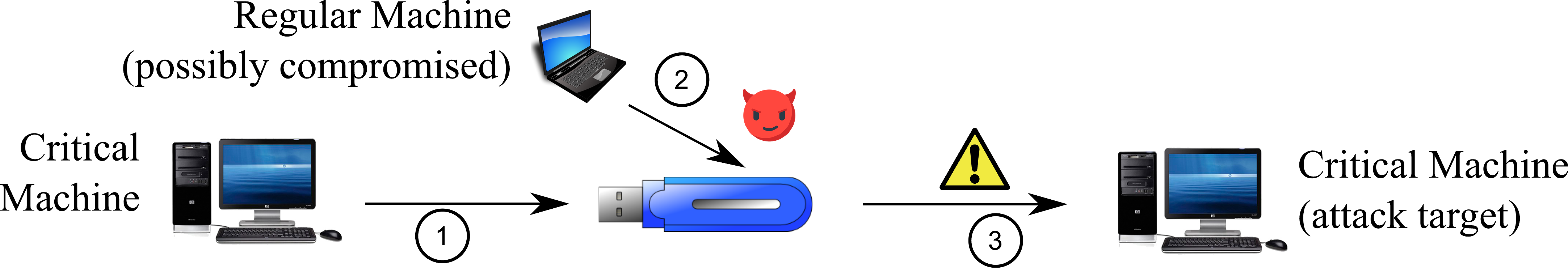}
	\caption{A promiscuous use of a RSD.}
	\label{fig:usecasesimplified}
\end{figure*}

About the promiscuous use of a RSD, consider the file copy scenario that is depicted in Fig.~\ref{fig:usecasesimplified}: \begin{inparaenum}[(1)]	
	\item a critical machine (e.g. a development workstation) writes some
	data into the RSD, e.g. a software update to be transferred into a critical machine,
	\item the RSD is plugged into a, possibly compromised, regular
	machine, which can infect the software update, add other malicious files in the
	RSD, or tamper with the firmware of the RSD, and
	\item the RSD is plugged into the critical machine (e.g. a SCADA
	server) that is the destination of the file transfer and also the target
	of the attack.
\end{inparaenum} 

Our goals are to allow this kind of use while preventing (potentially malicious)
data or code originated from regular machines to spread into critical ones as
well as prevent the damage of critical machines by malicious firmware. 
To achieve the first goal, we rely on cryptographic integrity protection along with
the use of authenticated data structures for efficiency.
Our approach does not rely on malware signatures and is a strong obstacle to the spread of zero-day attacks,
even if RSDes are used promiscuously in critical and regular machines.
To achieve the second goal, we rely on an hardware-based ``captcha''.
The basic idea is that the authenticity of a real \emph{Human Interface Device} (HID) can be
easily checked by asking the user to use it.
Our hardware-based solution does not require any change to host systems
and, hence, it is easily deployable in real ICS environments.

%


The rest of this paper is organized as follows. In Section~\ref{sec:backsoa}, we
review the state of the art.
Section~\ref{sec:actreq} introduces actors involved in our solution and
formalises the requirements we intend to meet. In Sections~\ref{sec:secmodel}, we
describe the security model and the threat model on which we base our work.
Section~\ref{sec:arch} shows the architecture of the proposed solution.
Section~\ref{sec:secanalysis} provides a security analysis.
In Section~\ref{sec:extensions}, we modify our solution to provide additional security features.
Section~\ref{sec:applcons} discusses the applicability of our approach in ICS
environments. In Section~\ref{sec:implementation}, we present a prototype of the
proposed solution and report informal feedback from experts. Section~\ref{sec:wrapup} draws the conclusions.

\section{State of the Art}\label{sec:backsoa}

To mitigate the risk for critical systems to be infected by a malware, an
antivirus can be adopted and properly configured to scan the data stored in the
USB thumb drive before any access.

Most commercial antiviruses perform detection based on a database of known
malware signatures. This approach has some drawbacks: it cannot detect zero-days
attacks, it needs regular signatures updates to keep its effectiveness,  its
performances depend on the size of the data to be protected, and it cannot
protect from generic tampering. 
Further, the class of attacks that leverage on the modification of the firmware (BadUSB attacks) makes
regular antivirus largely ineffective since they use capability this kind of
devices are allowed to use.
However, it may detect or block further malware actions occurring after the BadUSB attack.

The primary countermeasure proposed against BadUSB  was to \emph{protect} USB
devices by a firmware authentication feature that limits the firmwares that can
be uploaded into a device to those signed by the device vendor (see for
example~\cite{ironkeyfrmprt}).
However, this approach protects devices but not hosts, which are secured only if
they are forcibly limited to interact with just protected devices. Further, this
strongly limits the choice of USB storage devices and it is insecure, if the
limitation is delegated to error-prone user behaviour.

GoodUSB~\cite{tian2015defending} is a software solution that aims at protecting
the host against BadUSB attacks. When a new USB device is attached, a message is
shown to the user, which must declare his/her expectation about the
functionalities of the device. The user has to make a decision, that means there
is the possibility to incur in a human mistake or a deliberate malicious human behaviour.
A similar approach is adopted by~\cite{kang2017usbwall,loe2016sandusb}. USBlock~\cite{neuner2018usblock} consider 
the timing of USB traffic similarly to certain the intrusion detection approaches for 
IP networks.

There are a number of products on the market that specifically address security
for RSD (e.g., see~\cite{bitlockerwindows}) and USB thumb
drives (e.g., see~\cite{secureusblist}). These are mostly focused on
confidentiality, which, however, is not our primary objective. In these cases,
support to integrity is on a file basis or on a block basis, and there is no
integrity protection for the whole storage: an attacker can delete selected
portions of data and also revert part of them to a previously saved version.
Further, all solutions imply some form of authentication, usually
password-based, but once the user is authenticated, full access to data is
allowed, and a malware can easily infect the stored files.
Further, the adoption of a password as a protection impacts the usability in
term of ease to allow different people to use the device, i.e., the possibility to pass
the device from hand to hand.

\subsection{Integrity}\label{sec:integrity-soa}

Concerning techniques for checking the integrity of data, a large body of work
is known in literature. 

Protecting information by means of integrity in a scenario where exist different
type of machines (i.e. regular and critical) recalls the well known Biba integrity
model~\cite{biba1977integrity},
which describes a set of access control rules that can be used to protect the
integrity of certain data. In the Biba model, each element is associated with an
integrity level. The rules of this model deny any flow of information from lower
levels to higher levels and can be summarised with the statement ``no read down,
no write up''. This model is implemented in recent versions of the Windows
operating system~\cite{russinovich2009windows}. However, any form of access control on a filesystem
must be performed by a trusted operating system, while we want an USB thumb
drive to be promiscuously usable even on untrusted machines.

Many integrity approaches rely on robust cryptographic hash
functions~\cite{rogaway2004cryptographic}. When the dataset to be protected is
large, using hash functions is inefficient. In fact, for each change, even small
ones, the hash of the whole dataset have to be re-computed. Also, to check the
authenticity of a small part of the dataset, the hash of the whole dataset
should be checked. \emph{Authenticated Data Structures} (ADS) allow a user to
efficiently update a cryptographic hash of a large dataset when just a small
part of the dataset is changed. For an ADS, the hash of the whole dataset is
called \emph{root hash} or \emph{basis}. They also allow a user to efficiently
check the integrity of a small subset of data by only comparing against the root
hash an \emph{integrity proof} of size $O( \log n)$ with $n$ the size of the
data. Supposing that only the root hash is known to be genuine, it is possible to
check the integrity of a small subset of data, efficiently.

Widely-known ADSes are \emph{Merkle Hash Trees}
(MHT)~\cite{merkle1988digital} and authenticated skip
lists~\cite{goodrich2000efficient}. MHTs are balanced search trees
where leaves contains a cryptographic hash of the data and each
internal node contains a cryptographic hash of a concatenation of the
hashes stored in the children. For MHTs, the proof for a leaf $l$ is made of the 
hashes stored in the sibling of the nodes that are in the path from $l$ to the root.

For these data
structures, updates and checks are performed in logarithmic time with respect to
the size of the dataset, which is comparable to the efficiency of many indexes
for databases and filesystems. For this reason, MHTes or other ADSes have been
used in commercial, free, or research products. For example, MHTes were used for
securing filesystems (see for example, ~\cite{li2004secure,stefanov2012iris}).
Authenticated data structures were also adopted to authenticate relational database operations~\cite{devanbu2003authentic}. 
The problem of efficiently using ADSes with regular DBMS was studied in~\cite{miklau2005implementing,di2007authenticated,ppp-qrfccqr-10}.

\section{Actors and Requirements}\label{sec:actreq}

The actors of our solution are machines, humans, and USB devices. USB devices can be USB HIDes (for simplicity we consider only mice and keyboards) or RSDes.
Machines are divided in \emph{critical}
and \emph{non-critical} (or \emph{regular}). The set of critical (non-critical) machines is
the critical (non-critical/regular) \emph{realm}.
We consider critical realm as the part of the system that requires special
protection against malware generated in the non-critical realm and spread by
means of RSDes, which include thumb drives. 

%
In the USB protocol, the newly attached device declares its \emph{type} to the host
(i.e., if it is a keyboard, mouse, or RSD).
As for the protocol, a device is allowed to act only according to the
type it stated.
%

We consider two different possible ways to interact with machines:
\emph{data-flows} and \emph{inputs}.
We define data-flows as data transferred from machine to machine realised by means of read and write operations on RSDes.
We define inputs as data generated by device that allege to be HIDes.
We have a \emph{malicious data-flow} when the flow is from a regular machine to a critical machine.
We do not consider a data-flow as malicious when data pass through a regular machine to a critical machine with the mediation of a special component of the architecture called Gatekeeper (see Section~\ref{sec:arch}).
We have a \emph{malicious input} when the input is generated by a RSD that states to be an HID and, hence, the input itself was not generated by an interaction between a human and the HID.


In the following, we list the requirements that we have considered in the design of our solution.

\begin{description}

	\item [Discernment (Ds).]  The solution should prevent malicious data-flows and
malicious inputs from reaching  while allowing non-malicious ones.

	\item[Full Integrity (FI).] The solution should be able to detect malicious
data-flows. In other words, the solution has to detect all kinds of integrity violations,
comprising deletions and restoration of previous versions of files or parts of
them.


\item[Timeliness (T).] Malicious data-flows and malicious inputs should be detected
before they reach critical machines. 

\item[Interoperability (I).] The solution should be usable in conjunction with the
existing systems and software suites (SCADA, HMI, harsh laptops, development
environments, inventory management, etc.), without requiring any invasive
change to those products.

\item[Usability (U).] The solution should preserve the convenience and high
usability perceived by users when using RSDes and USB HIDes.

\item[Efficiency (E).] The solution should not introduce asymptotic complexity
overhead. Since operations for regular non-protected storage
technologies run in at most $O(\log n)$ time,
where $n$ is the amount of data stored, we mandate our solution cannot increase this 
complexity. For HID, we accept only constant time operations. 

%

	\item[Resiliency to Human Misbehaviour (R).] The solution should be not vulnerable to human
mistakes or intentional misbehaviour, hence, it has not to be based on any decision made by humans.
		
	\item[Determinism (Dt).] The solution should be regarded deterministic for any
practical purpose, that is, the probability that each single attack to be
successful should be so small to make any brute-force attack not viable.


\end{description}

\section{Security and Threat Model}\label{sec:secmodel}

We model an ICS as a set of machines (e.g. notebooks, workstations, SCADA,
embedded systems, etc) equipped with USB ports.
%
In our model we assume that only RSDes and HIDes can be used as USB devices.
Different USB devices are not allowed. 
RSDes and HIDes can be used promiscuously in both critical and non-critical realm.
We define a RSD as \emph{malicious } when it states to be an HID.

All data generated in the critical realm are considered trusted.
Data flows from regular to critical realm are forbidden and
allowed only using a \emph{gatekeeper} (see Section~\ref{subsec:gatekeeper}). All other
information flows should be allowed. This ideal setting conforms to the Biba
integrity model~\cite{biba1977integrity} with just two integrity levels, where
the rule ``no read down, no write up'' applies and the critical and regular
realms are the higher and lower integrity level respectively.

Physical keyboards and mice are considered  trusted and cannot be source of infections.
RSD devices are considered non-trusted due to the possibility to change the
firmware so that the device can act in a malicious way, i.e in a way that aim at damaging the system. 

In our model, we consider an attack to be 
\begin{inparaenum}[(1)]
	\item any write operation performed by a regular machine on something that is supposed to be read by a critical machine comprising addition, deletion and changes to data, metadata and directory structure, and
	\item any input generated by malicious RSDes that reaches a critical machine.
\end{inparaenum}

\section{Architecture}\label{sec:arch}

In this section, we describe the architecture of our solution. We
equip each critical machine with an hardware, that we call \emph{\U},
intended to be connected to one of the USB ports of the host. \U\ is itself provided with a
USB port where other USB devices can be connected. The idea is that \U\ should always be in the middle between
any USB device that is intend to be connected with a critical machine
and the critical machine itself. For this reason, we assume that \U\ cannot
be unplugged from the machine and each critical machine has all
available USB ports either protected by \U\ or disabled. 

\begin{figure*}
	\centering
	\includegraphics[width=1\columnwidth]{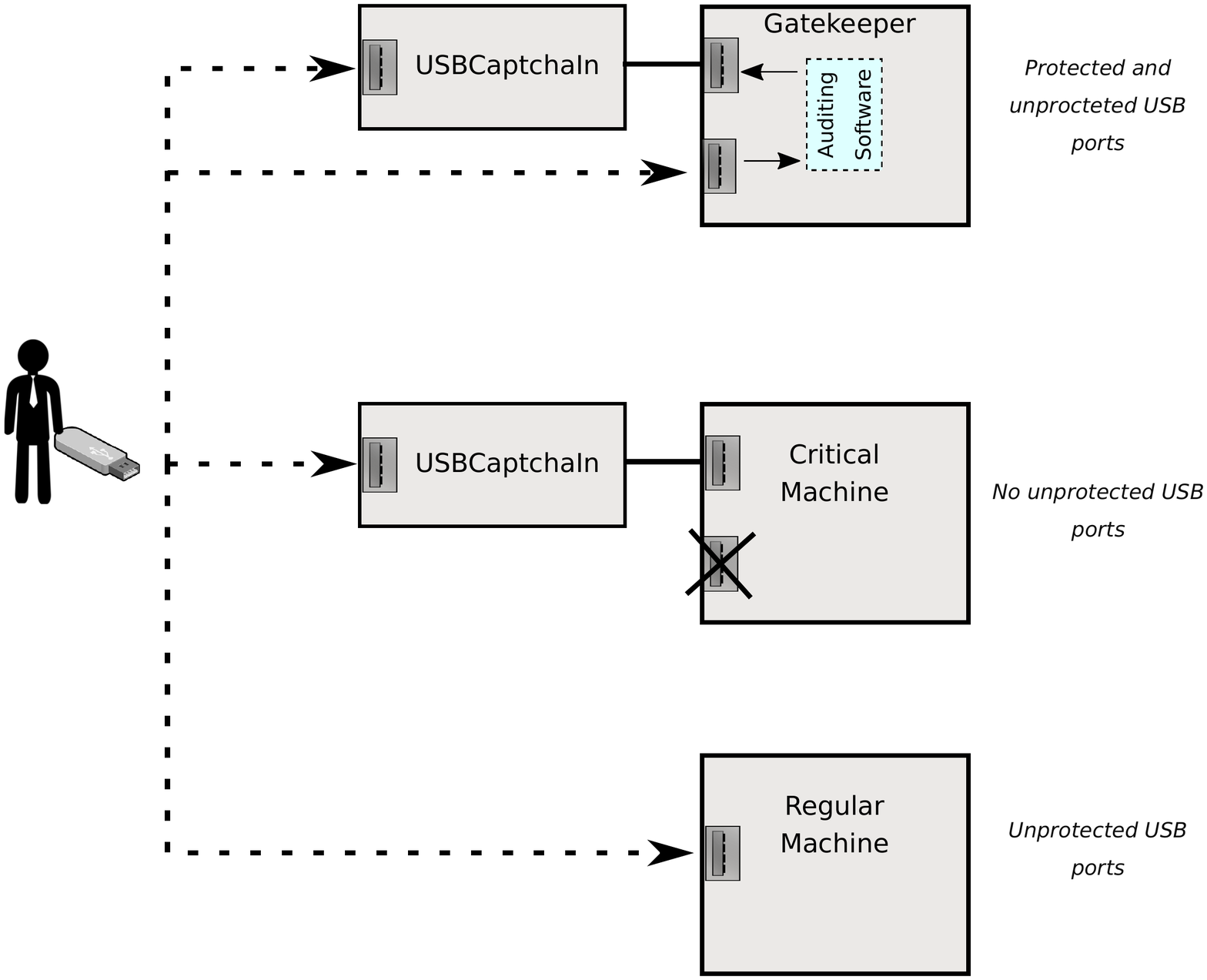}
	\caption{Components of the proposed architecture.}
	\label{fig:archcomponents}
\end{figure*}

 In our approach,
regular machines are not equipped with any specific hardware or
software. The general architecture is depicted in
Figure~\ref{fig:archcomponents}. We also consider a special
machine called \emph{gatekeeper} that allows exceptional data transfer
from regular to critical machines and whose details are described in
Section~\ref{subsec:gatekeeper}. We now focus on the internal
architecture of \U.

\subsection{\U}\label{sec:usbhwdevice}

\begin{figure*}
	\centering
	\includegraphics[width=1\columnwidth]{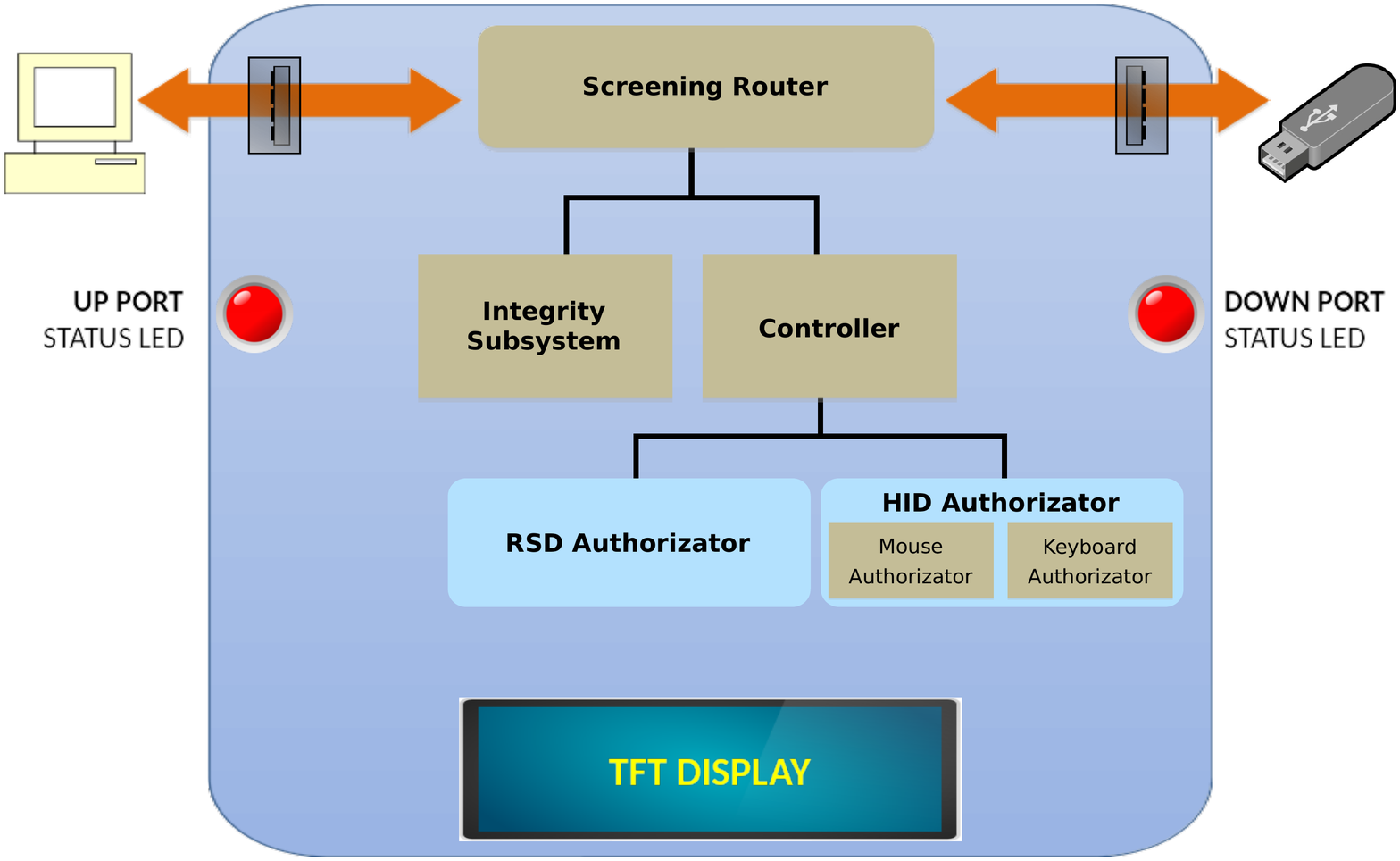}
	\caption{Components of the USBCaptchaIn device.}
	\label{fig:usbcaptchainArch}
\end{figure*}

The internal details of \U\ are shown in Figure~\ref{fig:usbcaptchainArch}.
\U\ is equipped with two USB ports: one \emph{upstream port} and one
\emph{downstream port} (also called \emph{up/down ports}). The up port of \U\ is
connected to one of the USB ports of a critical machine that we intend
to protect. All USB devices that a user intends to attach to the
critical machine should be attached to a down port. 
For the sake of simplicity, we present the architecture of \U\
with only one down port. The introduction of additional down ports is
a straightforward extension of the approach proposed in this paper.

Near the up port, there is an orange LED that indicates that the host
is powering \U\ and it is correctly working. Near the down port, there
is a status red/green LED. The LED blinks green when one of the
attached devices, possibly through a hub, is undergoing the
authorisation procedure, turns fixed green when all devices are
successfully authorized and can interact with the host, and turns
blinking red if the authorization procedure of one of the attached
devices failed too many times and the device must be disconnected. %
Messages required for user interaction are presented to the user on a
2.1" TFT display which is embedded in \U. %

In \U, a software intercepts, inspects, redirects, and injects USB
traffic between up and down ports.
When a user attaches the device to \U, the interaction between device
and \U\ follows the USB protocol: an \emph{enumeration} phase is
performed, during which  the device is supposed to provide specific
information about itself. \U\ listens the capabilities declared by the
device and records all information provided during enumeration. If the
device declares to be a RSD, \U\ behaves as described in
Section~\ref{subsec:intsysmod}. If it declares to be a HID, \U\
behaves as described in Section~\ref{subsec:act}. 
In both cases, the software realises the rules described in our security model.
The actual operation depends on the state of the
\emph{authorisation procedure} described in the above mentioned sections.

\U\ encompasses the following components.

\begin{itemize}

	\item [\textbf{Screening Router.}] The \emph{screening router} is in
	charge to route, USB packets to and from the
	ports, the controller, and the integrity subsystem. Its action depends on
	the authorization status of the device as explained in
	Sections~\ref{subsec:intsysmod} and~\ref{subsec:act}, which may end up in filtering out certain USB traffic. 
	It also performs standard USB
	enumeration, when a new device is connected.

	\item [\textbf{Integrity subsystem.}] The \emph{integrity subsystem} handles read and write
	operations related to RSDes and executes additional actions to
	guarantee that the data flows constraints of our security
	model are met (see Section~\ref{sec:secmodel}). The integrity
	techniques adopted are detailed in Section~\ref{subsec:intsysmod}.

	
	
	\item [\textbf{Controller.}] The controller orchestrates the display,
	the status leds, and the authorizators to realise the authorization
	procedure and human-machine interaction described in
	Section~\ref{subsec:intsysmod} and~\ref{subsec:act}. It also 	
	 reconfigures the screening router
	depending on the result of the authorisation procedure.
	
	\item [\textbf{Authorizators.}] The \emph{authorizators} manage the
	authorisation process of RSDes and supported HIDes types. 
	The correct authorizator is chosen by the
	controller according to what is declared by the device.
	The RSD Authorizator 
	performs some consistency checks (see
	Section~\ref{subsec:intsysmod}), while each HID Authorizator generates
	challenge codes for the user, keep track of the number of attempts,
	and check the correctness of the submitted code (see
	Section~\ref{subsec:act}).

\end{itemize}

\subsection{Integrity Subsystem and RSD Authorization}\label{subsec:intsysmod}

The integrity subsystem is in charge to ensure that data flows through
RSDes comply to the constraints described in
Section~\ref{sec:secmodel}. Essentially, any data transfer from
regular machines to critical machines should be blocked (exceptions
are handled as described in Section~\ref{subsec:gatekeeper}). To
efficiently perform this task and fulfil the Requirement~E, we base our solution on ADSes
(see Section~\ref{sec:integrity-soa}).

Each RSD has a \emph{secure partition} and an \emph{ADS partition}. In
the secure partition, we store data to be protected. In the ADS
partition, we store an ADS over the data do be protected plus some
additional cryptographic information. The state of the ADS is tightly
coupled with the content of the secure partition.

RSDes store sequentially-numbered equal-sized blocks of bytes. In the
USB protocol read and write operations issued by the host are at block
level. The integrity subsystem intercepts these operations and
inhibits those targeting blocks outside the secure partition.
Operations that target the secure partition are performed along with
additional tasks: reading encompasses integrity checks based on the
ADS and writing encompasses update of the ADS. The root-hash of the
ADS is kept signed in the ADS partition and used during the integrity
checks. In the following, we provide the details of our solution.

In our approach, the following kinds of operations can be identified.

\begin{itemize}
	
	\item[Protected-Read.] Protected-reads are read operations performed
	by the critical machine on data stored in the secure partitions of a RSD.
	Protected-reads are mediated by \U, which checks the integrity
	of the read data. If data is recognised as \emph{genuine}, the data
	is reported to the critical machine as result of the read. If data is
	recognised as \emph{tampered}, the read operation is blocked and an
	error is communicated to the critical machine. As detailed below, this fulfil Requirements~Ds, T, R and~Dt and partially FI. 
		
	\item[Protected-Write.] Protected-writes are the write operations
	performed by the critical machine on data stored in the secure
	partitions of a RSD. Protected-writes are mediated by \U, which
	additionally updates the ADS and the signature of the root hash.
	During the ADS update some integrity checks are performed. If they
	fail, an error is communicated to the critical machine.
	This contributes to fulfilment of Requirements Ds, T, R, and partially FI.

	\item[Illegal-Write.] An illegal-write is a write operation performed  on a
	secure partition by a regular machine. 
	The data changed by an illegal-write are always recognised
	by the subsequent protected-reads as tampered.
	
	\item[Plain-Read.] Plain-reads are normal read operations performed by regular
	machines when reading any part of a RSD comprising the secure partition.
	
\end{itemize}

Secure and ADS partitions are realised as regular partitions on the
RSD. Before the Integrity Subsystem starts to mediate the interaction between 
host and RSD, the RSD authorizator performs the following checks that guarantee the safety of read/write
operations performed by the Integrity Subsystem. 

\begin{enumerate}

\item It reads the partition table and check its compliance with its
standard format to exclude attacks at this level.

\item It identifies secure and ADS partitions. If they are not
present, this procedure is aborted and the RSD is not authorised.

\item It checks the correctness of format and size of the ADS
partition.

\item If the above actions are successful, it configures the screening
router to pass all host read/write requests to the Integrity
Subsystem. Further, it asks the Integrity Subsystem to initialise
itself for handling the identified secure and ADS partitions (see
below).

\end{enumerate}

Before describing the Integrity Subsystem we need to introduce some concepts.
Within each partition, we adopt the common approach of identifying blocks by a
numbering, assigning zero to the first block of that partition. The size of the
partitions are defined at the moment of their creation. Creation of partitions is handled by a
specific software that can create only partitions with \textit{empty state},
which are always recognised as genuine by \U. This software just resizes existing
partitions (like standard partitioning tools do) and create secure and ADS
partitions with the empty state. This procedure can be performed on regular machines without affecting security, since partition creation does not imply any data flow in the sense described in Sections~\ref{sec:actreq} and~\ref{sec:secmodel}.



Each \U\ device keeps, in local storage, its own \emph{private key} and a
corresponding \emph{certificate} signed by a unique
\emph{Certification Authortiy (CA)}, whose public key is also stored.
We denote by $U$ an instance of \U\ and by  $\mbox{CERT}(U)$ its
certificate.
Given a certain state of the secure partition $Z$, the hash of its
current content, i.e. the root-hash of the ADS, is denoted by $h(Z)$
and its signature is denoted by $\mbox{sign}_U(h(Z))$. The ADS
partition contains special fields named \emph{signature} and
\emph{last writer certificate}, that stores $\mbox{sign}_U(h(Z))$ and
$\mbox{CERT}(U)$, respectively. where $U$ is the last \U\ that wrote
in that RSD.

During initialisation of the Integrity Subsystem, right after the
authorisation of the RSD, the last writer certificate is read and
verified against the certificate of CA. The root-hash and its
signature are read and verified against the last writer certificate.
If any of these steps fails, the RSD is blocked and no operations are
allowed on it, otherwise the root-hash is considered trusted.

When the critical machine asks to read a block $b$ from $Z$ through $
U $, $U$ retrieves the proof of $b$ from the ADS (see Section~\ref{sec:backsoa}). If it is consistent
with the current trusted root-hash, the content of $b$ is deemed to be
genuine and is passed to the critical machine.
When the critical machine asks to update $Z$ through $ U $, the ADS
(comprising root-hash) and the signature should also be updated. This fulfils Requirements~Ds, T, R, and~Dt as far as access to RSDes is concerned. 

To fulfil Requirement~E, \U\
has a caching mechanism: it performs update of the ADS partition only
when there are no write operations pending, with a timer triggering the
actual write, similarly to what regular operating systems do. 
The last writer certificate is updated the first
time it is changed.



This approach does not completely fulfil Requirement FI. In fact, it
cryptographically detects all kinds of tampering except the full
reversion of both secure and ADS partitions to an older genuine
version. A version of our approach that completely fulfil Requirement FI is presented in Section~\ref{sec:extensions}. A
detailed security analysis is provided in
Section~\ref{sec:secanalysis}.

%

%



\begin{figure*}
	\centering
	\includegraphics[width=1\columnwidth]{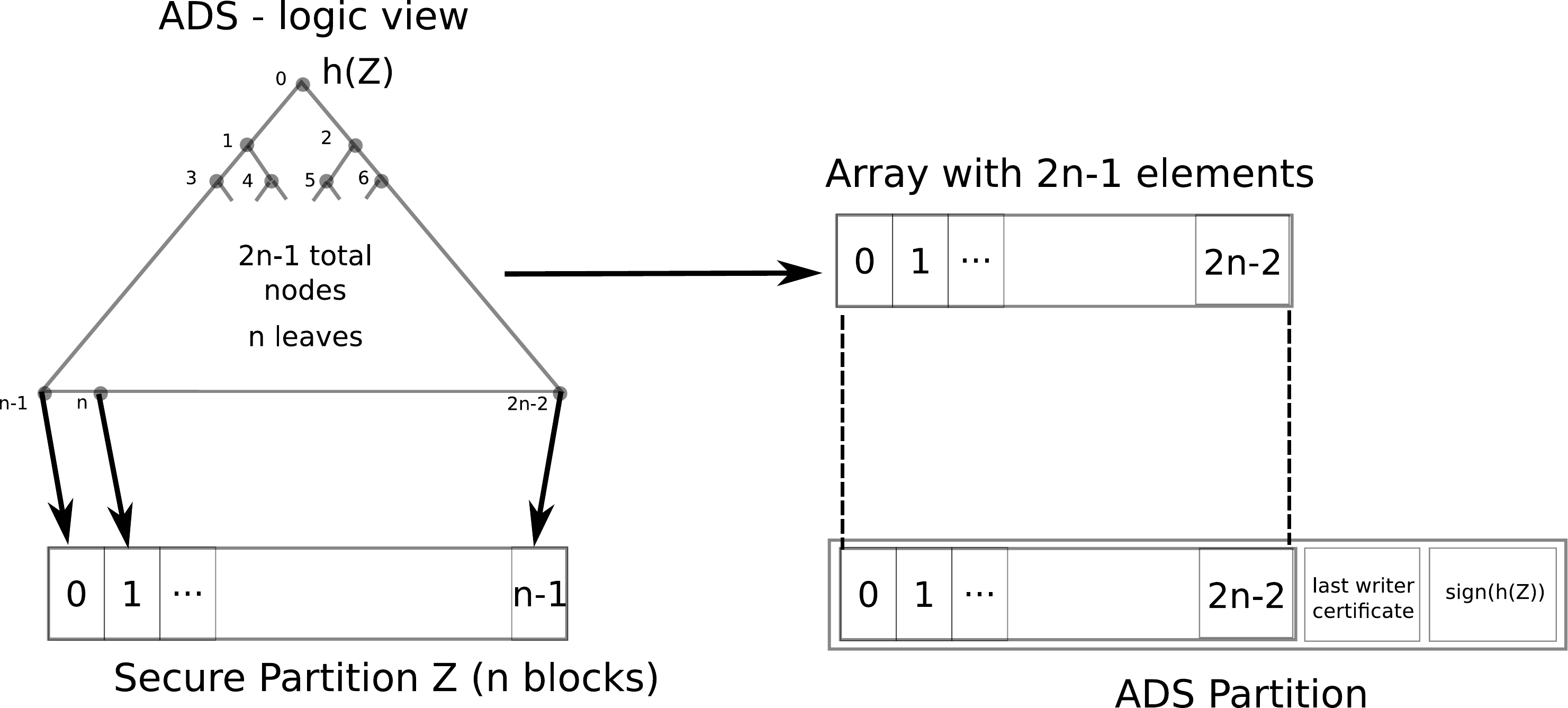}
	\caption{Relationship between secure partition, ADS, array, and ADS partition.}
	\label{fig:secadsrelationship}
\end{figure*}

We now describe the representation of the ADS in the ADS partition and its relationship with
the secure partition (see Figure~\ref{fig:secadsrelationship}).

For the sake of simplicity, we assume the secure partition contains $n$
blocks, where $n$ is a power of 2. In this case, our ADS has $ n $ leaves,
in one-to-one correspondence with the blocks of the secure partition,
and is  a complete binary Markle Hash Tree (see
Section~\ref{sec:integrity-soa}). From the properties of binary trees,
the total number of nodes (comprising internal ones) is $ 2n-1 $. Each
leaf of the ADS is the hash of the content of the corresponding block
of the secure partition. We represent the ADS with an array of $ 2n-1
$ elements (one for each node) following a sequential
representation. We assume the nodes of the ADS to be numbered
from $ 0 $ to $ 2n-2 $, from the root to the leaves following a
breath-first search order. According to this numbering scheme, each
node $ v $ has $ 2v+1 $ as left child and $ 2v+2 $ as right child. We
represent the ADS as an array whose elements correspond to the nodes
of the ADS according to the above defined numbering. The array 
stores only the hash, since the relationship between nodes are
implied by the above mentioned rules. The ADS partition
additionally stores signature and last write certificate, whose size
is fixed. Let $m$ be the size of the cryptographic hash and $B$ the
size of the blocks in bytes. The size of the representation of the ADS
is $(2n-1)m$ while the size of the secure partition is $nB$. It turns
out that, for large $n$,  the size of the ADS is $2m/B$ the size of
the secure partition. For example, for $B=4096$ (which is a typical
size for disk I/O) and $m=32$ (like for sha256), the ADS introduce a
storage overhead of just about 1.6\%.

\subsection{HID Authorisation}\label{subsec:act}

In this section, we describe the HID authorisation process and interaction between \U, the user, and
the HID (i.e., a keyboard or a mouse), that occurs when the latter is plugged into
the down port.

The
authorisation process is based on a physical interaction between human
beings and the HID just connected.
After the enumeration phase, \U\ starts the authorisation process asking to the user
to input, by means of the HID itself, a randomly-generated challenge code.

\begin{figure*}
	\centering
	\includegraphics[width=1\columnwidth]{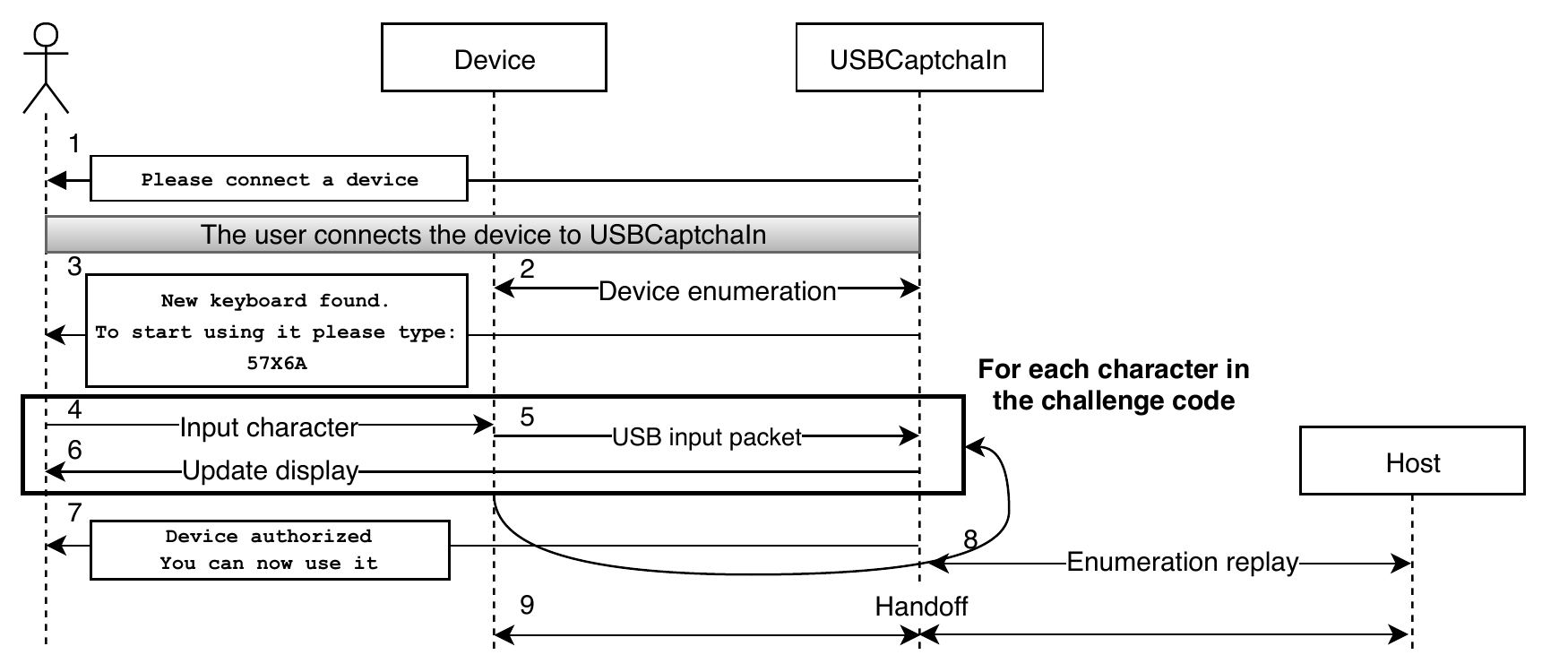}
	\caption{Interactions and messages among user, device, USBCaptchaIn, and host for HID authorization.}	
	\label{fig:behaviourHID}
\end{figure*}

The interaction for an HID that declares to be a keyboard is
summarised in Figure~\ref{fig:behaviourHID}. In this case, the
challenge code is a 5-characters alphanumeric string, which is
communicated to the user by showing on the display a proper message
with the code (Step 3 in Figure~\ref{fig:behaviourHID}). The user
types the \emph{elements} of the challenge code
(Figure~\ref{fig:behaviourHID}, Step 4), which for a keybord are
characters. Each typed character is sent by the device to \U\
according to the normal USB protocol for HID
(Figure~\ref{fig:behaviourHID}, Step 5). The authorizator checks that
each received element matches with the corresponding element of the
challenge code. If it matches, \U\ provides a visual feedback to the
user (Figure~\ref{fig:behaviourHID}, Step 6). For a
keyboard we colour green the corresponding correctly typed element on the display.
Then, \U\ loops expecting further elements and checking them until the
challenge code is finished.

If the challenge inputs are correctly inserted, \U\ notifies the user that the
new device has been authorised and that it can be used
(Figure~\ref{fig:behaviourHID}, Step 7).

At the same time, \U\ impersonates the device toward the host by
performing a ``replay'' of the enumeration that was previously
recorded (Figure~\ref{fig:behaviourHID}, Step 8). Then \U\ configures
the screening router to short-circuit (logically) the up and down
ports (Figure~\ref{fig:behaviourHID}, Step 9).

\U\ keeps monitoring the exchanged packets between the host and the device in
order to recognise hardware or logic \emph{detach/re-attach} operations and then
trigger a new authorisation procedure, when needed.

If any of the received elements does not match the corresponding element of the challenge code, the authorization is started over again showing a message that, for a keyboard, is like the following
\begin{verbatim}
Wrong code - try again.
To start using the keyboard pls type:
7E5N3
\end{verbatim}
The authorisation process can be tried (Figure~\ref{fig:behaviourHID}, steps 3-6) for a maximum of three times.
After three wrong attempts, \U\ definitely blocks the device, ignoring any more input from it, and shows a message on the display, as follows:
\begin{verbatim}
*** Authorization failed ***
Device claims to be a keyboard.
Is it true? Is the device malicious?
To check it again, re-attach it.
\end{verbatim}

To try again, the user has to detach and re-attach the device. This
physical interaction makes a completely automatic brute-force attack
impossible (see Section~\ref{sec:secanalysis} for further details).
Clearly, this is also the message displayed when a device maliciously
declares to be a keyboard and tries a repeated guess attack.
This approach fulfil Requirements~Ds and~T.

If the device is a mouse, \U\ asks the user to move a pointer on the display to
draw a line between two randomly-placed targets, for 3 times. The procedure was
selected to obtain a good compromise between security and usability.

\begin{figure*}
	\centering
	\includegraphics[width=1\columnwidth]{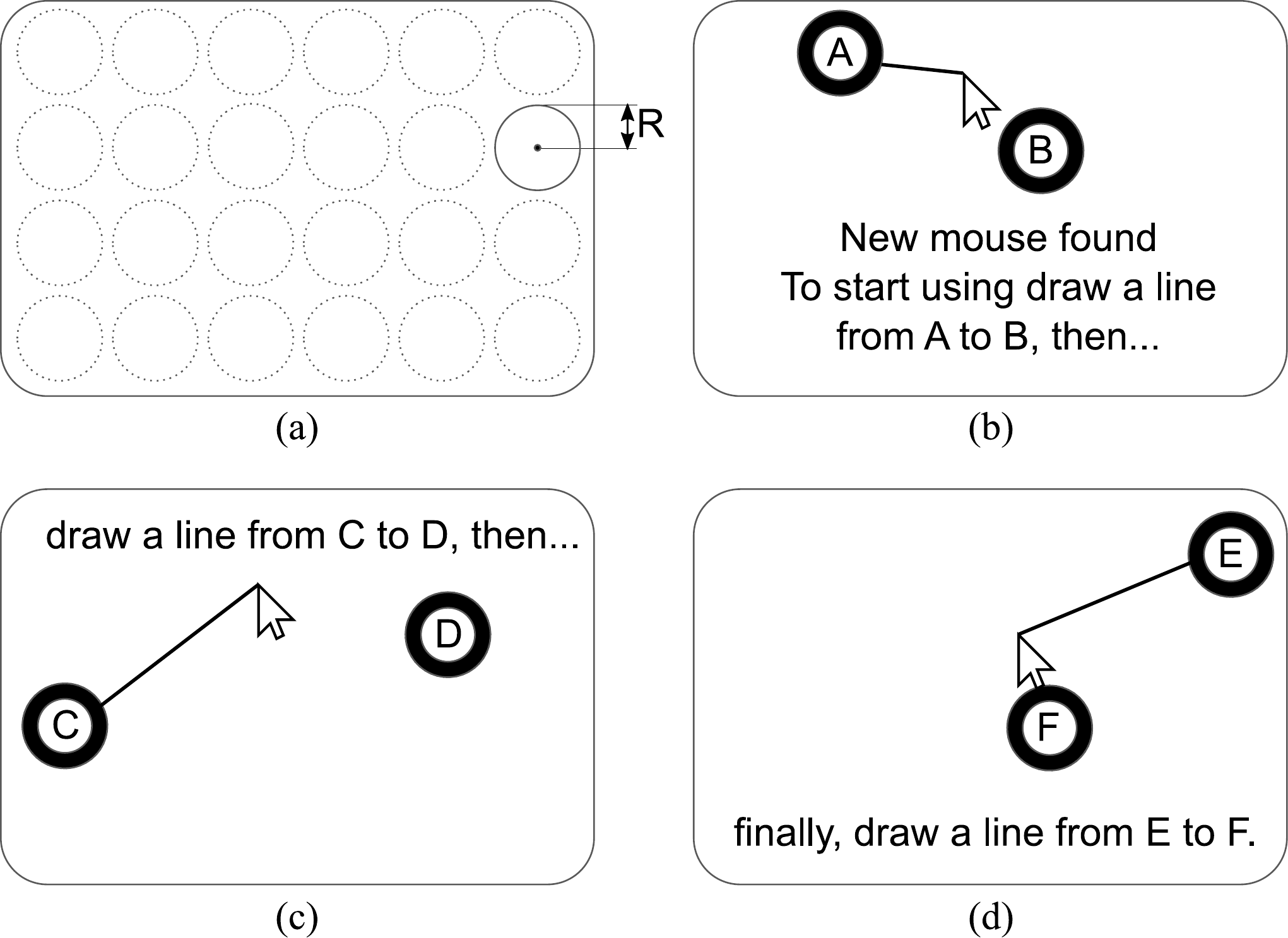}
	\caption{The mouse authorisation method. (a) Allowed target positions and
		radius $R$. (b,c,d) The three steps of the authorization: in (a) just two rows
		are available for second target positioning, in (b) and (c) three rows are
		available.}
	\label{fig:mouseAuth}
\end{figure*}

In this case, each element of the challenge is a pair of points of the \U\ display,
and the input of the user matches the element of the challenge if and only if
the distance between the points of the challenge and the clicked ones is below a
certain \emph{radius}. The radius is chosen so that 24 non-overlapping targets
can fit in the display (see Figure~\ref{fig:mouseAuth}).

Note that, instruction messages for the user are adaptively placed on
the display on an unused zone, after the first point was chosen, to
allow higher freedom in target positioning and make the required
interaction harder to guess for a malicious RSD that pretends to be a
mouse. See Section~\ref{sec:secanalysis} for a security analysis.

Once the HID is authorised, \U\ allows the HID to interact directly
with the host, without adding any overhead to the
USB packets exchanged by them, complying with Requirement~E.

\subsection{Gatekeeper}\label{subsec:gatekeeper}

The above described architecture ensures high level of security and
usability but does not support the relevant use case of bringing new
legitimate data or software into the critical realm, like for example manuals
or firmwares. They may be available in the regular realm and the first time they enter the 
critical realm we need to performs accurate audit.
The role of the gatekeeper is to ensure that these checks are performed in accordance with
the policy of the organisation. 

The gatekeeper realises the	``complete mediation'' security principle~\cite{ames1983security}. It plays the same role of firewalls in networking and of the security reference monitor for operating systems.
Practically, it is a dedicated machine that \begin{inparaenum}[(i)]
\item has some USB ports non-protected by \U\ from which any file can
be read, \item runs a specific software that performs thorough audits
on the read data, and \item writes the audited data on a secure
partition on a RSD plugged into \U, which in turn is connected to the
gatekeeper on a different port.
\end{inparaenum}
Clearly, the security of the software in the gatekeeper and of the whole gatekeeper machine 
is paramount.

The gatekeeper is the single point where data flow policies must be
configured and are enforced. While the actual policies may depend on
the specific industrial context, they should cover (a) authentication
of the operator that requires the transfer, (b) authorization of the
transfer according to the policy of the organization, and (c) logging
of the details of the operations performed to allow subsequent
auditing. Our gatekeeping approach enables to scale-up the security
level that an organization can achieve enabling the implementation of
arbitrarily thorough analysis of the data to be transferred and
arbitrarily complex workflow to obtain the authorisation, which may
involve human decision.

%
%
%

\section{Security Analysis}\label{sec:secanalysis}

The intent of this section is to explicit a number of assumptions and to show that, under those assumptions, with the adoption of the proposed solution, the critical realm cannot be compromised.
We recall that, protection from denial of service, i.e., making data
not accessible, is not among the objectives of the proposed solution.


%
%


We consider the following assumptions:

\begin{enumerate}[(I)]
	\item \label{assump:nosecbug} \U\ does not have security bugs and it is not compromised,
	\item \label{assump:throughrd} all HIDes and RSDes communicate with the host through \U,
	\item\label{assump:gk} the gatekeeper effectively checks (and possibly blocks) data
flowing from the regular realm toward the critical realm,
	\item \label{assump:nonetwork} critical machines cannot communicate with
regular machines by means of other then RSDes and trusted console operators,

	\item \label{assump:pki} attackers cannot obtain a private key (which
	we assume to be generated within \U\ at production time and
	never exported) or force the CA to sign a new certificate and CA is not
	compromised,
	
	\item \label{assump:cryptook} the adopted cryptographic primitives have no security flaws, and
	\item \label{assump:nounplug} an attacker cannot unplug \U\ from the host. 

\end{enumerate}

Our approach is not vulnerable to human mistakes or intentional misbehaviour. In fact, we do not ask to the
user to take any decision. Integrity checks process is totally transparent to
the user. During the authorisation process of HIDes, the user does not have to
accept device features or choose in a list of allowed device classes. Actually,
not even a malicious user can force \U\ to authorise a device that claims to be
an HID but offers no physical means to the user to provide the requested
challenge code.
By the above considerations, we can argue that our approach fulfil Requirement~R.

By Assumption~\ref{assump:nosecbug}, \U\ cannot be compromised by inserting RSDes that 
contains malformed data (e.g., a malformed partition table or ADS partition). 

By Assumption~\ref{assump:nonetwork}, the only vector of attack are RSDes plugged
into critical machines. By Assumptions~\ref{assump:throughrd}
and~\ref{assump:nounplug}, all communications between USB devices and critical
machines are mediated by \U.

By Assumption~\ref{assump:nosecbug}, there is no way to bypass the authorisation
process of HIDes than guessing and brute-force attack against the
\emph{challenge code}. %
While Assumption~\ref{assump:nosecbug} is a demanding one, honoring it for a dedicated hardware is
surely easier than honoring it for a software which runs on the host. In fact,
the security of \U\ is based on the security of a reasonably small and stable
amount of software, while for a host-based solution all software running on the
machine should be trusted unless proper mandatory access
control configurations are in place, which may not be feasible in many
environments. Further, certification procedures are much easier for an small embedded system
whose elements do not change.

The authorisation process of HIDes is extremely well protected against
guessing/brute-force attacks. For keyboard authorisation, the challenge code
consists of $ 5 $ characters each ranging within $ 26 $ letters plus $ 10 $
digits. The probability for a malicious device to correctly guess a random
challenge in three attempts is $3\ over\ 36^5$ (i.e., 1 over about 20 millions).
For mouse authorisation, each challenge code consists of $ 3 $ pairs of points.
The first point of each pair ranges within $ 24 $ possible positions. The second
point ranges within the unused positions that remain after the text message is
placed: $ 11 $ for the first elements (Figure~\ref{fig:mouseAuth}b) and
$ 17 $ for the second and third element (Figures~\ref{fig:mouseAuth}c
and~\ref{fig:mouseAuth}d). %
The probability of a successful attack in $ 3 $ attempts is $3\ over\
24^3\cdot11\cdot17^2$ (i.e., 1 over about 14 millions).
The above considerations show that our approach meet Requirement~Dt as far as HID authorisation is concerned. 

It is worth noting that a malicious device has no clue about the success or
failure of each attempt and after three failed attempts the human intervention
is required to gain more attempts. Devices recognised as RSDes cannot maliciously mimic
HID behaviour: once a device is authorised as non-HID, any attempt to mimic HID
behaviour requires a re-enumeration that can be triggered by a logic detach signal,
which in turn triggers a new authorisation process by \U.

\U\ cannot prevent a \emph{malicious HID} (i.e., a keyboard or
a mouse containing malicious code embedded in the firmware), to
actually allow the user to enter the challenge code. Other proposals,
like for example~\cite{tian2015defending}, have similar limitations. However,
solutions like~\cite{mulliner2018usblock} analyse the timing characteristics of 
the USB traffic which may detect a malicious HID. Nothing prevents to integrate 
into \U\ a similar approach.

About the integrity checks process, critical machines can read only data stored in
secure partitions. By Assumption~\ref{assump:gk}, no malicious file from the
regular realm is admitted in the critical realm. Hence, the only remaining
possibility for an attack is trying to make critical machines to read tampered
data from a secure partition.

Since RSDes are completely untrusted, the attacker (e.g., a
compromised regular machine) can freely tamper with any data stored in
the RSD (see Section~\ref{sec:secmodel}). This includes data stored in
the secure partition, and all data stored ADS partition, namely, ADS,
signature, and last writer certificate. Let us consider a tampering of
a secure partition. Since the result of any protected-read operation
is checked against the signed hash, through the proof derived by the ADS
in the ADS partition, the tampering of data stored in the secure
partition is easily detected.

Now, let us consider an attacker who replaces the content of the
secure and ADS partition with an old version of them. In this case,
the check of the signed hash with the result of a protected-read
matches. As mentioned in Section~\ref{subsec:intsysmod} we do not
completely fulfil Requirement FI. See Section~\ref{sec:extensions} for
a description of the improvement that we propose to the approach
described in Section~\ref{sec:arch} to fulfil Requirement FI completely.

The attacker can try to avoid detection by tampering also the ADS partition. 
An attack to the ADS, that does not change the root hash, requires to find a
collision for the hash function on which the ADS is based, which is against
Assumption~\ref{assump:cryptook}. On the other hand, to change the root hash the
attacker should be able to violate the signature. However, this attack
is ruled out by  Assumption~\ref{assump:cryptook} and, by
Assumption~\ref{assump:pki}. The attacker cannot get private key from \U\ for Assumption~\ref{assump:nosecbug}.

Tampering only with the signature, the last writer certificate, or the ADS 
ends up in a false positive. In fact, in those cases, while data contained in the secure
partition may be genuine, the integrity system has no way to prove it, hence, it
behaves as if the secure partition was corrupted denying any access to it. We recall that
ensuring data availability is not within the objective of our solution. %

If a protected-write operation is partially executed, for example because the
RSD is unplugged abruptly, the data in the secure partition, the ADS, the signature and the last writer certificate
may not be written or may be partially written. In this case,
a protected-read, detects a tampering.

\section{Ensuring Full Integrity and Providing Additional
	Functionalities}\label{sec:extensions}

As noted in Sections~\ref{subsec:intsysmod} and~\ref{sec:secanalysis}, the
solution presented till now, does not completely fulfil Requirement~FI. In fact,
reversion to a previous version of the secure parition, along with its consistent ADS parition, 
is not detected as
tampering. This may be regarded as an acceptable behaviour or not depending on
the context. In this section, we modify our solution so that Requirement~FI is
completely fulfilled. The changes that we propose are not free. They introduce
additional complexity, and in a certain sense, additional costs. This is the
reason why they are presented here as an improvement and not in the main design,
leaving open the possibility to adopt the limited version in situations in which
partial fulfilment of Requirement FI is acceptable. The changes described in
this section also offer the opportunity to provide additional functionalities
that may be desirable.

\begin{figure*}
	\centering
	\includegraphics[width=1\columnwidth]{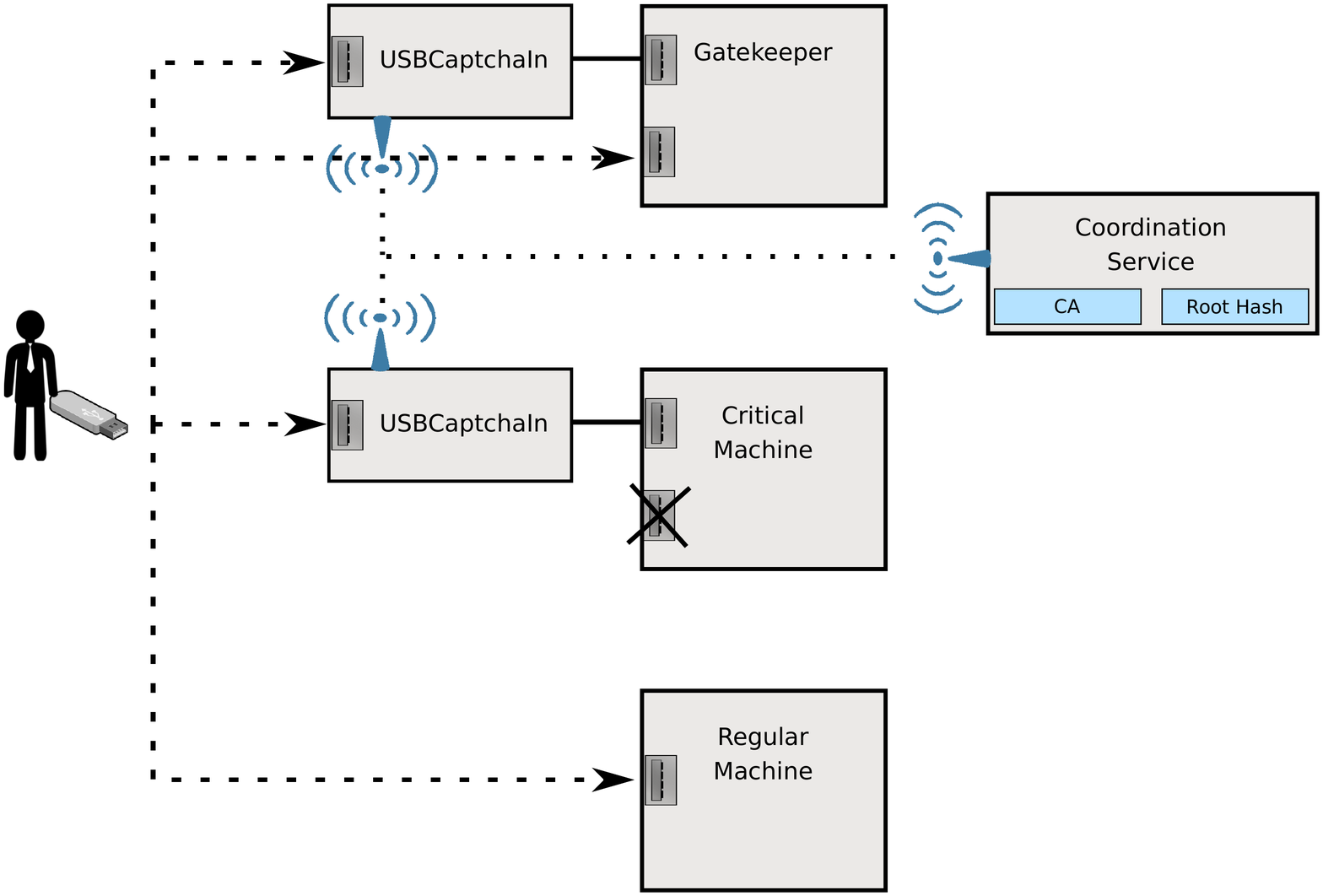}
	\caption{How the Coordination Service fits the architecture described in Section~\ref{sec:arch}.}
	\label{fig:archcomponents-extensions}
\end{figure*}

The new general architecture is shown in
Figure~\ref{fig:archcomponents-extensions}. We introduce a \emph{coordination
	service} (CS), whose functionality is to store key-value pairs in which each key
is associated with one RSD (for example, the USB device identifier may be used
for this purpose) and the value is the current root-hash of the ADS
(and in turn of the current state of the secure partition). To connect with CS, a communication
channel is needed. The bandwidth requirement for it is very small: only the RSD
key and the cryptographic hash (the final root-hash) for each update of the
secure partition should be sent. To support it, even a low bandwidth GPRS
connection may be enough. We note that the availability of CS and of the
connection might be an issue. In fact, if CS is not available or reachable, RSDes
cannot be used. However, nothing prevents the adoption of standard
high-availability approaches. Clearly, the communication
between \U\ and CS must be properly protected using standard technology (e.g.,
TLS~\cite{dierks2008transport}) and CS must be secured and authenticated.

If we accept the presence of a (low-bandwidth) communication channel, we can
exploit it to provide an additional remote administration/monitoring
functionality, which may be enabled on-demand under the control of CS. To
realise this, we equip \U\ with the capability to create a VPN toward CS, then a
data connection between \U\ and the critical machine can be created in two ways. If the
critical machine supports fully fledged USB protocol, \U\ may present itself during enumeration with
an additional functionality of USB network card. Otherwise, a physical cable may
connect \U\ with a physical network interface of the critical machine.

This architecture may also support a certificate revocation procedure. In fact, CS may 
periodically provide to all \U\ devices a certificate revocation
list and possibly distribute new certificates. 

Additional easy to implement functionalities are logging and monitoring of RSD activity and 
capability of administratively disable the use of RSDes on certain critical machines.

\section{Applicability Considerations}\label{sec:applcons}

In the following, we discuss the applicability of our approach in real ICS
environments. %



\paragraph{RSD Data Integrity Protection}
Contrary to
current suggested best practices~\cite{stouffer2011guide}, with our approach,
users are allowed to use the same RSD with any machine without relaying on
passwords and without the need to install specific software. It is enough to
equip critical machines with \U. Protections offered by the integrity subsystem  are
largely transparent to users, addressing Requirement~U (see Section~\ref{sec:actreq}). When an RSD is plugged into a \U, only the secure
partition is accessible to the critical machine. When an RSD is plugged into a
regular machine, secure and ADS partitions should not be touched.  To avoid
unintentional tampering of them, the secure partition can be logically write-protected using
capabilities supported by certain filesystems (e.g., NTFS) or partition tables
(e.g., GPT). In this cases, the secure partition can be mounted read-only on
regular machines. Note that, if the above solutions cannot be adopted (for
example for FAT-formatted drives with MBR partition tables), automatic
read-write mounting of a secure partition may happen on a regular machine, which
may end up in a detection of tampering due to automatic changes (e.g., update of
mounted-bit flag, auto-defragmentation, etc.). To avoid this issue, we slightly
change the behaviour described in Section~\ref{subsec:intsysmod}: we store the
content of the secure partition shifted of (at least) one block, so that, the
first block(s) are unused and zeroed. \U\ provides block numbers translations
on-the-fly during protected-read/write operations. In this way, the filesystem
cannot be recognised and mounting cannot occur on regular machines.

\paragraph{User Interaction with \U} We now discuss the interaction of the user with \U\ with respect to usability
(see Requirement~U). For Concerning RSDes, no user interaction is required to
authorise a RSD with well formatted secure and ADS partitions. HIDes
authorisation requires the users to use the device in a regular way. If the
device is a keyboard, they have to type. If it is a mouse, they have to click.
A keyboard or a mouse attached to \U\ can be used also to interact during the boot
process and to execute recovery procedures.



\paragraph{Deployment Impact}
It is easy to deploy \U\ in the critical realm since 
it does not need any driver on critical machines to work. \U\ is
totally independent from the operating system (OS) of the host it is connected
to. It can be connect to any
machine equipped with a USB port supporting standard USB protocol for keyboard, mouse and/or storage.

Our approach allows on organization to use any standard RSDes, promiscuously, complying with Requirement~I.

The gatekeeper provides the organization with great flexibility about the
security policies, but this means that they should be carefully designed and
possibly integrated with business or decision processes. For strict security
policies, traversing the gatekeeper may be costly, hence, it is advisable to
deploy a critical machine realizing a repository of commonly used files ready to
be used in the critical realm. 

The coordination service may be realised in several ways. The simplest
one, supporting only the full Requirement FI, is by ZooKeeper~\cite{hunt2010zookeeper}, which also supports 
high availability out-of-the-box.

%

The key management is quite simple. 
The integrity subsystem requires only
the presence of an off-line certification authority which may already be present
in the organization for other purposes. Aspects related with certificate
revocations have been dealt with in Section~\ref{sec:extensions}.

%

We assumed RSDes and HIDes communicate with critical machines through \U.
This assumption can be easily guaranteed in several ways. For instance, a costly
approach in term of money and changes required for the deployment, can  be to
integrate \U\ directly into the hardware of critical machines. Instead, a
cheaper approach, for example, can be to leave in critical machines only one USB
port enabled and weld \U\ to that port. Adopting one of these approaches, it is
impossible to bypass the mediation
of \U\ without physical tampering.

\section{Prototypical Realisation and Feedbacks from Experts}\label{sec:implementation}

We realised two prototypes that implement the more important parts of
the architecture described in Section~\ref{sec:arch}. Our main
objective is to evaluate the usability of \U\ in practice. We showed
the prototypes to ICS security experts. In this section, we also
report their feedback.

%

We focused on assessing the fulfilment of the 
usability requirement for the authorisation process for HID devices and for accessing RSDes, that is for the use of the integrity system module.

\begin{figure*}
	\centering
	\includegraphics[width=1\columnwidth]{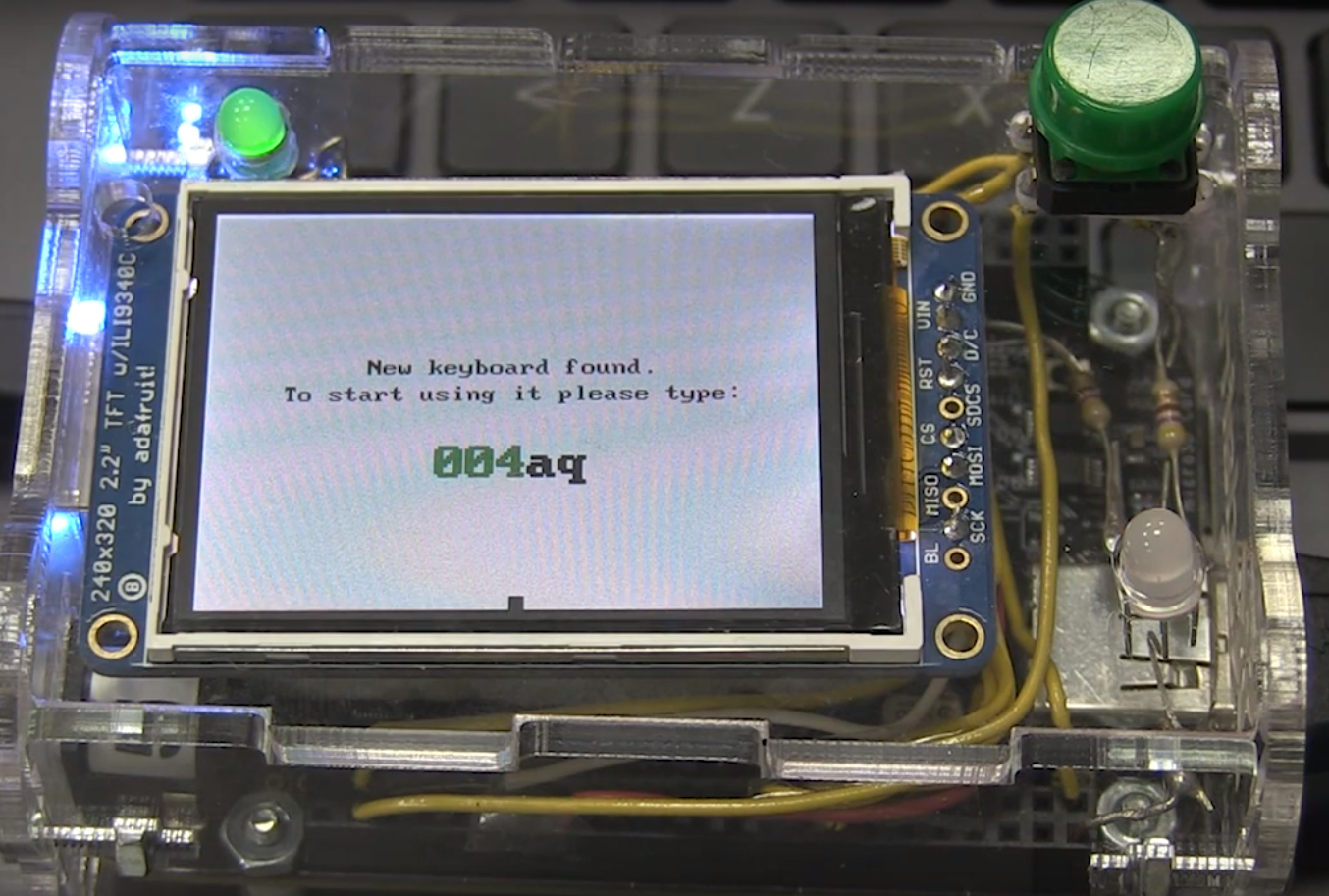}
	\caption{The hardware prototype for HID authorisation. The display shows a message during keyboard authorisation procedure.}
	\label{fig:usbchecinpic}
\end{figure*}

Our first prototype realises \U\ (see
Section~\ref{sec:usbhwdevice}) on a Beaglebone Black
board~\cite{bbb} on which a Linux kernel is running.
In this prototype, we realised the elements that are needed to test the HID authorisation 
process, namely, screening router, controller, and HID authorizators. 
The software running on \U\ is a customised
version of USBProxy~\cite{usbproxy}. The filter and the authorizator are
USBProxy plugins. The communication between the device
connected to a down port and the screening router is realised by
means of libUSB~\cite{libusb}, a library that supports the interaction
with generic USB devices, while the communication between
the host and the screening router is realised by means of gadgetFS,
a linux kernel module that allows the Beaglebone to act as a
client towards the host.
We packaged our prototype so that it can be used on real systems. The final result 
is shown in Figure~\ref{fig:usbchecinpic}.

Our second prototype mimics the functionalities of the integrity
system module (see Section~\ref{subsec:intsysmod}). It consists of a
software that, differently from our target design, is intended to be installed
in critical machines. However, it provides the same level of security
and a very similar user interaction.
In our simplified realisation, we intercept file system actions at the
system call level by using standard
open source drivers~\cite{fuse,dokany}. User files and directories are
stored in a regular directory of the RSD,
while the corresponding ADS is stored in an auxiliary
directory according to a proprietary format.
Files and directories can be accessed using the regular system calls of the operating 
system. Our software intercepts them and performs corresponding
operations on plain data and on the ADS using regular read and write
primitives of the kernel.
More details about this prototype can be found in~\cite{griscioli2016securing}.

We showed our prototypes to security experts that we met within
meetings of the Preemptive EU research
project~\cite{miciolino2017preemptive}. The feedback was very positive
for both of them. We recorded no complaints about usability.
One of the most appreciated aspects was the
possibility to realise them in independent hardware. Some of the
people we talked with, also suggested that \U\ could be physically
mounted within existing hardware whose USB ports should be protected, while keeping it isolated from the rest of the system for security. Some
have pointed out that promiscuously using RSDes formatted for use with
\U\ into other systems may raise annoying false positives due to users
accidentally manipulating integrity-protected data on a regular system.
However, this can be easily mitigated by exploiting support for soft
read-only protection of files (in the prototype) or partitions (in the
target design), which should avoid most of the accidental modifications, as described
in Section~\ref{sec:applcons}.

Concerning efficiency, on the HID authorisation side, once mice and
keyboards are authorised, the mediation of \U\ does not introduce any
noticeable delay for a user with respect to the case where the HIDes
are directly connected to the host. On the integrity protection side,
our experiments performed with our prototype have shown that the
additional overhead is negligible with respect to timings involved in
normal user interactions, like working on documents and opening
multimedia files. The overhead is mainly due to the fact that for each
operation on data a corresponding operation on the ADS must also be
performed. While for read operations this might be mitigated by proper
caching (which is automatically performed by the operating system in
our prototype), caching is not helpful for write operations. 
With the diffusion of USB 3.0 devices, this kind of overhead will
progressively be less and less important. On the other hand, the
limited computing power of a cheap board may severely limit the
transfer bandwidth to and from the RSD, with respect to the bandwidth supported
by USB 3.0. However, this aspect is more related with a market strategy trade-off 
and out of the scope of this paper.

\section{Conclusion}\label{sec:wrapup}

We presented an architecture based on a dedicated hardware 
for protecting critical machines in a ICS against malware spread by means of
USB devices.
\U\ realises a new approach to protect hosts, both against BadUSB attacks, in which
generic USB devices maliciously mimic the behaviour of HID
devices, and against malware embedded in data stored in RSDes.
The proposed approach proactively blocks attacks before they reach the target
and does not rely on decision of the users. It is highly compatible with already
deployed products, comprising embedded devices, like Programmable Logic
Controllers, Remote Terminal Units, etc.,  on which is often not feasible to install
new software.

The user experience is only slightly altered by the protection offered by our solution and the presence of \U\ does not impact USB devices
responsiveness. 

Our informal contacts with experts, to whom we have shown our prototypes, have reported that a solution like the one we described might be
accepted in a real ICS environment.

Further research work may encompass the study of supporting a Biba
security model with more than two security levels. This may be useful,
for example, to support a split of the critical realm in a \emph{production}
realm (highly critical) and a \emph{testing} realm (less critical).
Further classes of HIDes may also be  supported (e.g., touch tablet, track-ball, joystick, etc).
The integration in \U\ of a solution like the one described in~\cite{mulliner2018usblock}
may add a form of protection against real keyboards with
malicious firmware without affecting security and usability.

\section{Acknowledgements}
We would like to thank Marco Sacchetti and Diego Pennino for their contribution to the realisation of the prototypes.
This work was partially supported by the European Commission under project "Preemptive" (grant agreement n. 607093).

\bibliographystyle{abbrv}
\bibliography{bibliography}

\end{document}